\documentclass[11pt]{article}
\usepackage{amsmath,amssymb}

\newcommand{\ba}{\begin{alignat}{3}}
\newcommand{\e}{\epsilon}

\newcommand{\dl}{\delta}

\newcommand{\g}{\gamma}

\newcommand{\sg}{\sigma}

\newcommand{\pa}{\partial}

\newcommand{\om}{\omega}
\newcommand{\mc}{\mathcal}

\begin{document}

\begin{titlepage}
\begin{flushright}
%    IU-MSTP/ \\
%    25 August 2011
\end{flushright}
\begin{center}
  \vspace{3cm}
  {\bf \Large Super Virasoro Algebras From Chiral Supergravity}
  \\  \vspace{2cm}
  Yoshifumi Hyakutake
   \\ \vspace{1cm}
   {\it College of Science, Ibaraki University \\
   Bunkyo 1-1, Mito, Ibaraki 310-0062, Japan}
\end{center}

\vspace{2cm}
\begin{abstract}
 
\end{abstract}
In this note, we construct Noether charges for the chiral supergravity, which contains 
the Lorentz Chern-Simons term, by applying Wald's prescription to the vielbein formalism.
We investigate the AdS$_3$/CFT$_2$ correspondence by using the vielbein formalism.
The asymptotic symmetry group is carefully examined by taking into account the local Lorentz transformation,
and we construct super Virasoro algebras with central extensions from the chiral supergravity.
\end{titlepage}

\setlength{\baselineskip}{0.65cm}

%%%%%%%%%%%%%%%%%%%%%%%%%%%%%%%%%%%%%%%%%%%%%%%%%%%%%%%%%%%%%%%%%%%%%%%%%%%%%%%%%%%%%%%%%%%%%%%%%%

\section{Introduction}

%%%%%%%%%%%%%%%%%%%%%%%%%%%%%%%%%%%%%%%%%%%%%%%%%%%%%%%%%%%%%%%%%%%%%%%%%%%%%%%%%%%%%%%%%%%%%%%%%%

The three dimensional gravity with negative cosmological constant has been one of the 
interesting testing grounds to uncover quantum natures of gravity.
Especially the gauge/gravity correspondence has been investigated from various aspects for decades.

The vacuum solution of the three dimensional gravity with negative cosmological constant 
is described by global AdS$_3$ geometry\cite{Deser:1983nh}.
In 1986, Brown and Henneaux showed that the asymptotic symmetry group of the AdS$_3$ geometry consists of 
left and right Virasoro algebras, 
and they succeeded to evaluate the same central charges for both algebras\cite{Brown:1986nw}.
This is a prototype of the gauge/gravity correspondence, which was conjectured sophisticatedly
in the context of superstring theory\cite{Maldacena:1997re}.
The three dimensional theory also contains BTZ black hole solution which is found 
by Banados, Teitelboim and Zanelli\cite{Banados:1992wn,Banados:1992gq}.
And the entropy of the BTZ black hole is statistically explained by using the Cardy formula for 
the boundary CFT\cite{Strominger:1997eq}.
It is well known that the three dimensional gravity theory can also be described by 
the gauge Chern-Simons theory\cite{Achucarro:1987vz,Witten:1988hc}.
The Virasoro algebras can be derived by using this alternative formulation\cite{Banados:1994tn},
and the black hole entropy is statistically explained in ref.~\cite{Carlip:1994gy}.

There are many important works on the three dimensional gravity, but we just focus on
three kinds of generalizations on the Virasoro algebras at the boundary.
First one is to deal with the supergravity\cite{Coussaert:1993jp}. 
As expected, the asymptotic symmetry group enhances to 
super Virasoro algebras and the central charges can be
evaluated including fermionic sector\cite{Banados:1998pi,Henneaux:1999ib}.
Second one is to add chiral terms to the theory.
The three dimensional gravity with the gravitational or Lorentz Chern-Simons term is called topologically 
massive gravity (TMG)\cite{Deser:1982vy,Deser:1981wh}.
In this theory it has been studied that the central charges for left and right modes are 
asymmetric\cite{Maldacena:1997de}-\cite{Hotta:2008yq}.
Third one is to consider higher derivative corrections, such as $R^2$ terms.
In this case, central charges are modified 
by some conformal factors\cite{Saida:1999ec,Hotta:2008yq,Azeyanagi:2009wf}.

The purpose of this note is to consider the supergravity with negative cosmological constant
which contains the Lorentz Chern-Simons term.
The supergravity with the Lorentz Chern-Simons term, which is called 
the topologically massive supergravity (TMSG), 
is constructed by Deser and Kay\cite{Deser:1982sw},
and the cosmological constant is added to the TMSG by Deser (CTMSG)\cite{Deser:1982sv}.
There are two parameters in CTMSG : the cosmological constant $-\frac{2}{\ell^2}$ 
and the coefficient of the Lorentz Chern-Simons term $\beta$.
It is known that fluctuation around the AdS$_3$ geometry contains negative energy mode
for generic $\ell$ and $\beta$\cite{Li:2008dq}.
The exception occurs at the critical point $|\beta/\ell|=1$, and the theory is called 
chiral supergravity\cite{Li:2008dq,Becker:2009mk}.
Since we need stable AdS$_3$ background to explore the gauge/gravity correspondence, 
the chiral supergravity is investigated in this note.
We employ Wald's prescription to construct the Noether charge for the 
chiral theory\cite{Wald:1993nt}-\cite{Tachikawa:2006sz}.
Especially we formulate the chiral supergravity in the vielbein formalism\cite{Hyakutake:2012uv}.
The charges are covariant under the general coordinate transformation,
and it is possible to evaluate the asymmetric central charges for left and right modes explicitly.
As a result, super Virasoro algebras at the boundary are explicitly constructed, which are expected from
the viewpoint of AdS/CFT correspondence\cite{Becker:2009mk}.
The vielbein formalism is applicable to all supergravity theories\cite{Hyakutake:2012uv,Hyakutake:2014maa},
and this work will be useful to test the gauge/gravity correspondence in superstring theory and 
M-theory at quantum level\cite{Hanada:2013rga,Hyakutake:2013vwa}.

In section 2, we explain some basic properties of the CTMSG.
In section 3, we construct the current for the general coordinate transformation
and that for the local supersymmetry.
We review the asymptotic symmetry group of the AdS$_3$ in section 4.
The super Virasoro algebras for the chiral supergravity are constructed and 
the central charges for left and right movers are derived in section 5.
Section 6 is devoted to the conclusion and discussion.

%%%%%%%%%%%%%%%%%%%%%%%%%%%%%%%%%%%%%%%%%%%%%%%%%%%%%%%%%%%%%%%%%%%%%%%%%%%%%%%%%%%%%%%%%%%%%%%%%%

\section{Cosmologically Topologically Massive Supergravity}
\label{sec:EOM}

%%%%%%%%%%%%%%%%%%%%%%%%%%%%%%%%%%%%%%%%%%%%%%%%%%%%%%%%%%%%%%%%%%%%%%%%%%%%%%%%%%%%%%%%%%%%%%%%%%

The topologically massive supergravity (TMSG) is the three dimensional supergravity with
Lorentz Chern-Simons term which was constructed by Deser and Kay\cite{Deser:1982sw}.
Deser also generalized the theory by adding the cosmological constant (CTMSG)\cite{Deser:1982sv}.
In this section we review the equations of motion for the CTMSG.
Fields of the CTMSG consist of a vielbein $e^a{}_\mu$ and a Majorana gravitino $\psi_\mu$.
Here $\mu, \nu$ are used for space-time indices and $a,b = 0,1,2$ are for local Lorentz ones.
In this note we consider $\mathcal{N}=(1,0)$ CTMSG\footnote{If the sign of $\ell$ is flipped, 
we obtain $\mathcal{N}=(0,1)$ CTMSG. Although the bulk gravity has three dimensions,
by taking into account the AdS/CFT correspondence, 
we use the notation $\mathcal{N}=(1,0)$ in the boundary CFT.}, 
and the Lagrangian is given by
\ba
  \mathcal{L} &= \frac{e}{16\pi G_\text{N}} \Big\{ R + \frac{2}{\ell^2} 
  - \frac{1}{2} \overline{\psi_\rho} \g^{\mu\nu\rho} \psi_{\mu\nu} \notag
  \\
  &\quad\;
  + \frac{\beta}{2} \e^{\mu\nu\rho} \Big( \om_\mu{}^a{}_b \partial_\nu \om_\rho{}^b{}_a 
  + \frac{2}{3} \om_\mu{}^a{}_b \om_\nu{}^b{}_c \om_\rho{}^c{}_a \Big)
  - \frac{\beta}{2} \overline{D_\rho \psi_\sigma} \g^{\mu\nu} \g^{\rho\sigma} D_\mu \psi_\nu \Big\}.
  \label{eq:CSGLag}
\end{alignat}
Here $G_\text{N}$ and $-2/\ell^2$ are the gravitational constant and the negative cosmological one, respectively.
$\beta$ is a coefficient for the nonchiral part.
Since we evaluate physical quantities in the background of AdS$_3$ with $\psi_\mu = 0$ 
in later sections, below we consider the Lagrangian up to $\mathcal{O}(\psi^3)$.

In eq.~(\ref{eq:CSGLag}), two kinds of covariant derivatives are defined,
\begin{alignat}{3}
  D_\mu \psi_\nu &= \partial_\mu \psi_\nu + \frac{1}{4} \omega_{\mu ab} \g^{ab} \psi_\nu, \qquad
  \mathcal{D}_\mu \psi_\nu &= D_\mu \psi_\nu + \frac{1}{2\ell} \g_\mu \psi_\nu, \label{eq:covD}
\end{alignat}
and the field strength of the Majorana gravitino is given by 
$\psi_{\mu\nu} \equiv \mathcal{D}_\mu \psi_\nu - \mathcal{D}_\nu \psi_\mu$.
The gamma matrix in three dimensions satisfy the Clifford algebra $\{\gamma^a, \gamma^b \} = 2 \eta^{ab}$, 
and $\eta^{ab} = \text{diag}(-1,1,1)$. 
The gamma matrix with spacetime index is defined as $\g^\mu = e^\mu{}_a \g^a$,
and a completely antisymmetric tensor $\gamma^{\mu_1\cdots \mu_n}$ is defined 
so that a coefficient of each term becomes $1/n!$. 
$\gamma^{\mu\nu\rho} = \e^{\mu\nu\rho} \bf{1}$ is a completely antisymmetric tensor in three dimensions.

The spin connection is expressed in terms of the vielbein and the Majorana gravitino
by requiring $D_\mu \big( 2 e e^\mu{}_a e^\nu{}_b \big) = 
\frac{1}{4} e \overline{\psi_\rho} \gamma^{\rho\mu\nu} \g_{ab} \psi_\mu$.
After standard calculations, the explicit forms of the spin connection and its variation can be obtained as
\begin{alignat}{3}
  \om_{\rho ab} &= e^\mu{}_{[a} e^\nu{}_{b]} \Big(- e_{\rho c} \partial_{\mu} e^c{}_{\nu} 
  + e_{\mu c} \partial_\nu e^c{}_{\rho} - e_{\mu c} \partial_{\rho} e^c{}_\nu \notag
  \\
  &\qquad\qquad\quad\,
  + \frac{1}{4} \overline{\psi_\mu} \g_\rho \psi_\nu 
  - \frac{1}{4} \overline{\psi_\nu} \g_\mu \psi_\rho 
  + \frac{1}{4} \overline{\psi_\rho} \g_\mu \psi_\nu \Big), \label{eq:spin}
  \\[0.1cm]%%%%%%%%%%%%%%%%%%%%%%%%%%%%%%%%%%%%%%%%%
  \dl \om_{\rho ab} &= e^\mu{}_{[a} e^\nu{}_{b]} \Big( - e_{\rho c} D_\mu \dl e^c{}_\nu
  + e_{\mu c} D_\nu \dl e^c{}_\rho - e_{\mu c} D_\rho \dl e^c{}_\nu \notag
  \\
  &\qquad\qquad\quad\,
  + \frac{1}{2} \overline{\dl \psi_\mu} \g_\rho \psi_\nu 
  - \frac{1}{2} \overline{\dl \psi_\nu} \g_\mu \psi_\rho 
  + \frac{1}{2} \overline{\dl \psi_\rho} \g_\mu \psi_\nu \Big). \label{eq:varspin}
\end{alignat}
Then, up to $\mathcal{O}(\psi^2)$, the variation of the Lagrangian (\ref{eq:CSGLag}) becomes
\ba
  16 \pi G_\text{N} \dl \mathcal{L} 
  &= 2 e \Big\{ R^a{}_\mu - \frac{1}{2} e^a{}_\mu \Big(R + \frac{2}{\ell^2} \Big) \Big\} \dl e^\mu{}_a 
  - e \overline{\dl \psi_\rho} \gamma^{\rho\mu\nu} \psi_{\mu\nu} \notag
  \\
  &\quad\,
  + \frac{\beta}{2} \big( - e \e^{\rho\mu\nu} R^{ab}{}_{\mu\nu} \dl \om_{\rho ab}
  + 2 e \overline{\dl \psi_\sigma} \g^{ab} \g^{\rho\sigma} D_\rho D_a \psi_b \big) \label{eq:CSGvar1}
  \\
  &\quad\,
  + \pa_\mu \Big( 2 e e^\mu{}_a e^\nu{}_b \dl \om_\nu{}^{ab} 
  + e \overline{\psi_\nu} \gamma^{\mu\nu\rho} \dl \psi_\rho
  + \frac{\beta}{2} e \e^{\mu\nu\rho} \om_{\nu ab} \delta \om_\rho{}^{ab} 
  - \beta e \overline{\dl \psi_\nu} \g^{\rho\sigma} \g^{\mu\nu} D_\rho \psi_\sigma \Big). \notag
\end{alignat}
In the above calculation, we used $D_\mu \g^a = 0$ and $D_\gamma (e \g^{\mu\nu\rho}) = 0$.
Note that $D_\rho (e \g^{\rho\mu_1 \cdots \mu_n}) = \mathcal{O}(\psi^2)$ from eq.~(\ref{eq:spin}),
and we employed this relation to derive eq.~(\ref{eq:CSGvar1}).

Let us evaluate the first term in the second line in eq.~(\ref{eq:CSGvar1}).
The Riemann tensor in three dimensions is written in terms of the Ricci tensor and the scalar curvature as
\begin{alignat}{3}
  &R_{\mu\nu\rho\sigma} = g_{\mu\rho} R_{\nu\sigma} - g_{\mu\sigma} R_{\nu\rho}
  - g_{\nu\rho} R_{\mu\sigma} + g_{\nu\sigma} R_{\mu\rho}
  -\frac{1}{2} (g_{\mu\rho} g_{\nu\sigma} - g_{\mu\sigma} g_{\nu\rho}) R, \notag
  \\
  &\e^{\rho\mu\nu} R^{ab}{}_{\mu\nu} 
  = 2 \e^{\rho a\sg} R^b{}_\sg - 2 \e^{\rho b\sg} R^a{}_\sg - \e^{\rho ab} R. \label{eq:3dR}
\end{alignat}
By using eqs.~(\ref{eq:varspin}) and (\ref{eq:3dR}), the first term in the second line in eq.~(\ref{eq:CSGvar1})
is evaluated as
\begin{alignat}{3}
  - \frac{\beta}{2} e \e^{\rho\mu\nu} R^{ab}{}_{\mu\nu} \dl \om_{\rho ab} 
  &= 2 \beta e \e^{\mu\nu\rho} C_{a\rho}
  \Big( - D_{\mu} \dl e^a{}_{\nu} + \frac{1}{2} \overline{\dl \psi_\mu} \g^a \psi_\nu \Big). \label{eq:CSGvar2}
\end{alignat}
In this calculation we used
\begin{alignat}{3}
  C_{\mu\nu} &= R_{\mu\nu} - \frac{1}{4} g_{\mu\nu} R, \qquad
  \e^{\rho\mu\nu} R^{ab}{}_{\mu\nu} = 2 \e^{\rho a\sg} C^b{}_\sg - 2 \e^{\rho b\sg} C^a{}_\sg, \label{eq:Ctensor}
\end{alignat}
and we neglected terms of $\mathcal{O}(\psi^2)$.

Finally the variation of the Lagrangian (\ref{eq:CSGvar1}) is expressed as
\begin{alignat}{3}
  \delta \mathcal{L} &= 
  \frac{e}{16\pi G_\text{N}} \big( 2 G^a{}_\mu \delta e^\mu{}_a + \overline{\delta\psi_\rho} \Psi^\rho \big) 
  + \frac{1}{16\pi G_\text{N}} \partial_\mu \big( e \Theta^\mu(\dl) \big).
  \label{eq:CSGvariation}
\end{alignat}
In the above we defined
\begin{alignat}{3}
  \Theta^\mu(\dl) &= 2 e^\mu{}_a e^\nu{}_b \delta \omega_{\nu}{}^{ab} 
  + \bar{\psi}_\nu \g^{\mu\nu\rho} \delta \psi_\rho \notag
  \\
  &\quad\,
  + \frac{\beta}{2} \e^{\mu\nu\rho} \om_{\nu ab} \delta \om_\rho{}^{ab} 
  - 2 \beta \e^{\mu\nu\rho} C_{a\rho} \dl e^a{}_{\nu} 
  - \beta \overline{\dl \psi_{\nu}} \g^{\rho\sg} \g^{\mu\nu} D_\rho \psi_\sg, \label{eq:Theta}
\end{alignat}
and
\begin{alignat}{3}
  G{}^a{}_\mu &\equiv R^a{}_\mu - \frac{1}{2} e^a{}_\mu \Big( R + \frac{2}{\ell^2} \Big)
  + \beta \e^{a\nu\rho} e^b{}_\mu D_{\nu} C_{b\rho}, \notag
  \\
  \Psi^\rho &\equiv - \g^{\mu\nu\rho} \psi_{\mu\nu} + \beta \e^{\mu\nu\rho} C_{a\nu} \g^a \psi_\mu
  - \beta \g^{ab} \g^{\rho\sg} D_\sg D_a \psi_b. \label{eq:EOM}
\end{alignat}
The equations of motion for the CTMSG are given by $G^a{}_\mu = 0$ and $\Psi^\rho = 0$.

%%%%%%%%%%%%%%%%%%%%%%%%%%%%%%%%%%%%%%%%%%%%%%%%%%%%%%%%%%%%%%%%%%%%%%%%%%%%%%%%%%%%%%%%%%%%%%%%%%%

\section{Currents for the CTMSG}

%%%%%%%%%%%%%%%%%%%%%%%%%%%%%%%%%%%%%%%%%%%%%%%%%%%%%%%%%%%%%%%%%%%%%%%%%%%%%%%%%%%%%%%%%%%%%%%%%%%

The action of the CTMSG is invariant under the general coordinate transformation 
and the local supersymmetry.
In this section we will construct currents for these transformations via Wald's 
procedure\cite{Wald:1993nt,Iyer:1994ys}.

%%%%%%%%%%%%%%%%%%%%%%%%%%%%%%%%%%%%%%%%%%%%%%%%%%%%%%%%%%%%%%%%%%%%%%%%%%%%%%%%%%%%%%%%%%%%%%%%%%%
\subsection{Current for the General Coordinate Invariance}
%%%%%%%%%%%%%%%%%%%%%%%%%%%%%%%%%%%%%%%%%%%%%%%%%%%%%%%%%%%%%%%%%%%%%%%%%%%%%%%%%%%%%%%%%%%%%%%%%%%

Let us consider the general coordinate transformation $x'^\mu = x^\mu - \xi^\mu$.
The vielbein and the spin connection transform as vector fields, and these behave like
\ba
  &\dl_\xi e^a{}_\mu = \xi^\nu \pa_\nu e^a{}_\mu + \pa_\mu \xi^\nu e^a{}_\nu
  = D_\mu \xi^a - \xi^\nu \om_\nu{}^a{}_\mu, \notag
  \\
  &\dl_\xi \omega_\nu{}^{ab} = \xi^\rho \pa_\rho \om_\nu{}^{ab} + \pa_\nu \xi^\rho \om_\rho{}^{ab}
  = \xi^\rho R^{ab}{}_{\rho\nu} + D_\nu (\xi^\rho \om_\rho{}^{ab}). \label{eq:GRtr}
\end{alignat}
Below we apply Wald's procedure to construct the current for the general coordinate 
transformation\cite{Wald:1993nt,Iyer:1994ys}\footnote{Noether's procedure is generalized to 
the gravitational Chern-Simons term in ref.~\cite{Tachikawa:2006sz}}.

First, by imposing the equations of motion $G^a{}_\mu = 0$ and $\Psi^\rho = 0$,
the variation of the Lagrangian (\ref{eq:CSGvariation}) becomes
\begin{alignat}{3}
  \dl_\xi \mathcal{L} = \frac{1}{16\pi G_\text{N}} \pa_\mu (e \Theta^\mu(\xi)). \label{eq:EOMLvar}
\end{alignat}
And the explicit form of $e \Theta^\mu(\xi)$ up to $\mathcal{O}(\psi)$ is evaluated 
as\footnote{Although eq.~(\ref{eq:Theta}) is expressed up to $\mathcal{O}(\psi^3)$ for the general coordinate
transformation, we also need to know the correct equations of motion (\ref{eq:EOM}) up to $\mathcal{O}(\psi^3)$
to obtain fermionic bilinear terms of $Q_{\mu\nu}(\xi)$. 
Thus we evaluate $\Theta^\mu(\xi)$ up to $\mathcal{O}(\psi)$, which is enough to obtain the super Virasoro
algebras in section \ref{sec:superVirasoro}.}
\begin{alignat}{3}
  &e \Theta^\mu(\xi) \notag
  \\
  &= 2 e e^\mu{}_a e^\nu{}_b \delta_\xi \omega_{\nu}{}^{ab} 
  + \frac{\beta}{2} e \e^{\mu\nu\rho} \om_{\nu ab} \delta_\xi \om_\rho{}^{ab}
  - 2 \beta e \e^{\mu\nu\rho} C_{a\rho} \dl_\xi e^a{}_{\nu} \notag
  \\%%%%%%%%%%%%%%%%%%%%%%%%%%%%%%%%%%%%%%%%%%%%%%%%%
  &= 2e R^\mu{}_\nu \xi^\nu + 2 e e^\mu{}_a e^\nu{}_b D_\nu (\xi^\rho \om_\rho{}^{ab})
  - \frac{\beta}{2} e \e^{\mu\nu\rho} \om_{\nu ab} \xi^\sg R^{ab}{}_{\rho\sg}
  + \frac{\beta}{2} e \e^{\mu\nu\rho} \om_{\nu ab} D_\rho (\xi^\sg \om_\sg{}^{ab}) \notag
  \\
  &\quad\,
  - 2 \beta e \e^{\mu\nu\rho} C_{a\rho} D_\nu \xi^a
  + 2 \beta e \e^{\mu b\rho} C^a{}_\rho \xi^\sg \om_{\sg ab} \notag
  \\%%%%%%%%%%%%%%%%%%%%%%%%%%%%%%%%%%%%%%%%%%%%%%%%%%
  &= 2e G^\mu{}_\nu \xi^\nu + e \Big( R + \frac{2}{\ell^2} \Big) \xi^\mu
  + \pa_\nu \Big( 2 e e^\mu{}_a e^\nu{}_b \xi^\rho \om_\rho{}^{ab}
  - 2 \beta e \e^{\mu\nu\rho} C_{\rho\sg} \xi^\sg 
  - \frac{\beta}{2} e \e^{\mu\nu\rho} \om_{\rho ab} \om_\sg{}^{ab} \xi^\sg \Big) \notag
  \\
  &\quad\,
  - \frac{\beta}{2} e \e^{\mu\nu\rho} \om_{\nu ab} R^{ab}{}_{\rho\sg} \xi^\sg 
  - \frac{\beta}{4} e \e^{\mu\nu\rho} \om_{\sg ab} R^{ab}{}_{\nu\rho} \xi^\sg
  - \frac{\beta}{2} e \e^{\mu\nu\rho} \om_\nu{}^a{}_b \om_\rho{}^b{}_c \om_\sg{}^c{}_a \xi^\sg \notag
  \\[0.1cm]%%%%%%%%%%%%%%%%%%%%%%%%%%%%%%%%%%%%%%%%%%%%%%%%%%
  &= 2e G^\mu{}_\nu \xi^\nu + \xi^\mu \mathcal{L} + \pa_\nu \big( e Q^{\mu\nu}(\xi) \big). \label{eq:Thetaxi}
\end{alignat}
Here we defined the antisymmetric tensor,
\begin{alignat}{3}
  e Q^{\mu\nu}(\xi) &= 2 e e^\mu{}_a e^\nu{}_b \xi^\rho \om_\rho{}^{ab}
  - \frac{\beta}{2} e \e^{\mu\nu\rho} \big( 4 C_{\rho\sg} + \om_{\rho ab} \om_\sg{}^{ab} \big) \xi^\sg. \label{eq:Q}
\end{alignat}
In order to obtain the last line in eq.~(\ref{eq:Thetaxi}), we used the relation
$\epsilon^{\mu\nu\rho} A_{\nu\rho\sigma} \xi^\sigma = \frac{1}{3} \e^{\nu\rho\sg} A_{\nu\rho\sg} \xi^\mu$
for a completely antisymmetric tensor $A_{\mu\nu\rho}$.

Second, since the Lagrangian of the CTMSG is covariant under the general coordinate transformation,
its variation behaves as a scalar field like
\begin{alignat}{3}
  \dl_\xi \mathcal{L} = \pa_\mu \big( \xi^\mu \mathcal{L} \big). \label{eq:Lagvar}
\end{alignat}
Note that the Lorentz Chern-Simons term is invariant under the general coordinate transformation.

Subtracting eq.~(\ref{eq:Lagvar}) from eq.~(\ref{eq:EOMLvar}), we obtain the conservation law of the current. 
The current for the general coordinate invariance is expressed as
\ba
  e J^\mu(\xi) &= \frac{1}{16\pi G_\text{N}} \big\{ e \Theta^\mu (\xi) - 16\pi G_\text{N} \xi^\mu \mathcal{L} 
  + \pa_\nu \big( e \tilde{Q}^{\mu\nu}(\xi) \big) \big\} \notag
  \\
  &= \frac{1}{16\pi G_\text{N}} \pa_\nu \big( e Q^{\mu\nu}(\xi) + e \tilde{Q}^{\mu\nu}(\xi) \big). 
  \label{eq:SGcurrent}
\end{alignat}
Here the equation of motion $G^\mu{}_\nu = 0$ is used, 
and $\tilde{Q}^{\mu\nu}(\xi)$ is an antisymmetric tensor. 
According to the Wald's procedure, in order to make the Hamiltonian well defined, the variation
of $\tilde{Q}^{\mu\nu}(\xi)$ should become
\ba
  \dl \big( e \tilde{Q}^{\mu\nu}(\xi) \big) = 
  e \big( \xi^\mu \Theta^\nu(\dl) - \xi^\nu \Theta^\mu(\dl) \big). \label{eq:Qtvar}
\end{alignat}
Then the variation of the current is evaluated as
\begin{alignat}{3}
  \dl \big( e J^\mu(\xi) \big) 
  &= \frac{1}{16\pi G_\text{N}} \pa_\nu \big\{ \dl \big( e Q^{\mu\nu}(\xi) \big) 
  + e \big( \xi^\mu \Theta^\nu(\dl) - \xi^\nu \Theta^\mu(\dl) \big) \big\}. \label{eq:Jvar}
\end{alignat}
We will use this expression to derive the Virasoro algebras from the chiral supergravity
in section \ref{sec:superVirasoro}.

%%%%%%%%%%%%%%%%%%%%%%%%%%%%%%%%%%%%%%%%%%%%%%%%%%%%%%%%%%%%%%%%%%%%%%%%%%%%%%%%%%%%%%%%%%%%%%%%%%%
\subsection{Supercurrent}
%%%%%%%%%%%%%%%%%%%%%%%%%%%%%%%%%%%%%%%%%%%%%%%%%%%%%%%%%%%%%%%%%%%%%%%%%%%%%%%%%%%%%%%%%%%%%%%%%%%

Let us construct the supercurrent for the CTMSG.
Under the local supersymmetry transformation, the vielbein and the Majorana gravitino transform as
\ba
  \dl_\e e^a{}_\mu &= \overline{\e} \g^a \psi_\mu, \qquad
  \dl_\epsilon \psi_\mu &= 2 \mathcal{D}_\mu \e. \label{eq:susyvariation}
\end{alignat}
Here $\epsilon(x)$ represents a spacetime dependent parameter which belongs to the Majorana representation. 
From these, we see that the variation of the spin connection and that of the field strength of 
the Majorana gravitino become
\begin{alignat}{3}
  \dl_\e \om_{\rho ab} &= e^\mu{}_{[a} e^\nu{}_{b]} \Big( - \bar{\e} \g_\rho D_\mu \psi_\nu
  + \bar{\e} \g_\mu D_\nu \psi_\rho - \bar{\e} \g_\mu D_\rho \psi_\nu \notag
  \\
  &\qquad\qquad\quad\,
  - \frac{1}{2\ell} \bar{\e} \g_\mu \g_\rho \psi_\nu 
  + \frac{1}{2\ell} \bar{\e} \g_\nu \g_\mu \psi_\rho 
  - \frac{1}{2\ell} \bar{\e} \g_\rho \g_\mu \psi_\nu \Big), \notag
  \\
  \dl_\e \psi_{\mu\nu} &= \frac{1}{2} R_{ab\mu\nu} \g^{ab} \e + \frac{1}{\ell^2} \g_{\mu\nu} \e, \label{eq:svar}
\end{alignat}
up to $\mathcal{O}(\psi^2)$.

First, by imposing the equations of motion $G^a{}_\mu = 0$ and $\Psi^\rho = 0$,
the variation of the Lagrangian (\ref{eq:CSGvariation})
becomes
\begin{alignat}{3}
  \dl_\e \mathcal{L} = \frac{1}{16\pi G_\text{N}} \pa_\mu (e \Theta^\mu(\e)). \label{eq:EOMLvarsusy}
\end{alignat}
And the explicit form of $\Theta^\mu(\e)$ up to $\mathcal{O}(\psi^2)$ is evaluated as
\begin{alignat}{3}
  \Theta^\mu(\e) &= 2 e^\mu{}_a e^\nu{}_b \delta_\e \omega_{\nu}{}^{ab} 
  + \bar{\psi}_\nu \g^{\mu\nu\rho} \delta_\e \psi_\rho \notag
  \\
  &\quad\,
  + \frac{\beta}{2} \e^{\mu\nu\rho} \om_{\nu ab} \delta_\e \om_\rho{}^{ab} 
  - 2 \beta \e^{\mu\nu\rho} C_{a\rho} \dl_\e e^a{}_{\nu} 
  - \beta \overline{\dl_\e \psi_{\nu}} \g^{\rho\sg} \g^{\mu\nu} D_\rho \psi_\sg. \label{eq:Thetaep}
\end{alignat}

Next, by consulting the calculations in section \ref{sec:EOM}, 
the variation of the Lagrangian under the local supersymmetry is evaluated as
\ba
  16 \pi G_\text{N} \dl_\e \mathcal{L} 
  &= 2 e \Big\{ R^a{}_\mu - \frac{1}{2} e^a{}_\mu \Big(R + \frac{2}{\ell^2} \Big) \Big\} \dl_\e e^\mu{}_a
  - \frac{1}{2} e \overline{\dl_\e \psi_\rho} \gamma^{\rho\mu\nu} \psi_{\mu\nu}
  - \frac{1}{2} e \overline{\psi_\rho} \gamma^{\rho\mu\nu} \dl_\e \psi_{\mu\nu} \notag
  \\
  &\quad\,
  + 2 \beta e \e^{\mu\nu\rho} C_{a\rho}
  \Big( - D_{\mu} \dl_\e e^a{}_{\nu} + \frac{1}{2} \overline{\dl_\e \psi_{\mu}} \g^a \psi_{\nu} \Big) 
  - e \beta \overline{D_\rho \dl_\e \psi_\sigma} \g^{\mu\nu} \g^{\rho\sigma} D_\mu \psi_\nu \notag
  \\
  &\quad\,
  + \pa_\mu \Big( 2 e e^\mu{}_a e^\nu{}_b \dl_\e \om_\nu{}^{ab} 
  + \frac{\beta}{2} e \e^{\mu\nu\rho} \om_{\nu ab} \dl_\e \om_\rho{}^{ab} \Big). \label{eq:svarL}
\end{alignat}
In the above, terms of $\mathcal{O}(\psi^2)$ are neglected.
The second and third terms in the first line of eq.~(\ref{eq:svarL}) are deformed as
\begin{alignat}{3}
  &- \pa_\rho \big( e \bar{\e} \gamma^{\rho\mu\nu} \psi_{\mu\nu} \big)
  + e \bar{\e} \gamma^{\rho\mu\nu} \mathcal{D}_\rho \psi_{\mu\nu}
  + \frac{1}{4} e R_{ab\mu\nu} \bar{\e} \g^{ab} \gamma^{\rho\mu\nu} \psi_\rho
  + \frac{1}{2\ell^2} e \bar{\e} \g_{\mu\nu} \gamma^{\rho\mu\nu} \psi_\rho \notag
  \\
  &= - \pa_\mu \big( e \bar{\e} \gamma^{\mu\nu\rho} \psi_{\nu\rho} \big)
  + 2 e \Big\{ R^a{}_\mu - \frac{1}{2} e^a{}_\mu \Big(R + \frac{2}{\ell^2} \Big) \Big\} \bar{\e} \g^\mu \psi_a.
  \label{eq:svarSG}
\end{alignat}
And the second line of eq.~(\ref{eq:svarL}) is calculated like
\begin{alignat}{3}
  &- 2 \beta e \e^{\mu\nu\rho} C_{a\rho}
  \Big( \bar{\e} \g^a D_\mu \psi_\nu + \frac{1}{2\ell} \bar{\e} \g_\mu \g^a \psi_{\nu} \Big) \notag
  \\
  &\quad\,
  + \frac{\beta}{4} e R_{ab\rho\sigma} \bar{\e} \g^{ab} \g^{\mu\nu} \g^{\rho\sigma} D_\mu \psi_\nu
  + \frac{\beta}{\ell} e \overline{D_\rho \e} \g_\sigma \g^{\mu\nu} \g^{\rho\sigma} D_\mu \psi_\nu \notag
  \\%%%%%%%%%%%%%
  &= \frac{\beta}{\ell} e \e^{\mu\nu\rho} C_{a\rho} \bar{\e} \g_\nu \g^a \psi_{\mu} 
  + \frac{2\beta}{\ell} e \overline{D_\rho \e} \g^{\mu\nu\rho} D_\mu \psi_\nu \notag
  \\%%%%%%%%%%%%%
  &= \pa_\mu \Big( \frac{2\beta}{\ell} e \bar{\e} \g^{\mu\nu\rho} D_\nu \psi_\rho \Big). \label{eq:svarLCS}
\end{alignat}
In order to derive the above expressions, we noted $\g^{\mu\nu\rho} = \e^{\mu\nu\rho} {\bf 1}$,
$\g^{\mu\nu} = \e^{\mu\nu\rho} \g_\rho$, $\g^\mu = - \frac{1}{2} \e^{\mu\nu\rho} \g_{\nu\rho}$, and
used relations below.
\ba
  &\g^{\mu\nu} \g^{\rho\sg} = - 2 g^{\rho[\mu} \g^{\nu]\sg} + 2 g^{\sg[\mu} \g^{\nu]\rho}
  - 2 g^{\rho[\mu} g^{\nu]\sg}, \notag
  \\
  &\g_\sg \g^{\mu\nu} \g^{\rho\sg} = 2 g^{\rho[\mu} \g^{\nu]} + 2 \g^{[\mu} \g^{\nu]\rho}
  = 2 \g^{\mu\nu\rho},  \notag
  \\%%%%%%%%%%%%
  &- \e^{\mu\nu\rho} R_{a\rho} \bar{\e} \g^a D_\mu \psi_\nu = 
  - R_{a\rho} \bar{\e} \g^{\mu\nu\rho} \g^a D_\mu \psi_\nu = 
  - 2 \bar{\e} R^{[\mu}{}_\rho \g^{\nu]\rho} D_\mu \psi_\nu
  - R \bar{\e} \g^{\mu\nu} D_\mu \psi_\nu, \label{eq:gammarelation}
  \\%%%%%%%%%%%%
  &R_{ab\rho\sigma} \g^{ab} \g^{\mu\nu} \g^{\rho\sigma} 
  = 16 R^{[\mu}{}_{\rho} \g^{\nu]\rho} + 6 R \g^{\mu\nu}. \notag
\end{alignat}
Eventually the variation of the Lagrangian for the CTMSG (\ref{eq:svarL}) becomes
\begin{alignat}{3}
  16\pi G_\text{N} \delta_\epsilon \mathcal{L} &= 
  \partial_\mu \Big( 2 e e^\mu{}_a e^\nu{}_b \delta_\epsilon \omega_{\nu}{}^{ab} 
  - e \bar{\epsilon} \g^{\mu\nu\rho} \psi_{\nu\rho} 
  + \frac{\beta}{2} e \e^{\mu\nu\rho} \om_{\nu ab} \delta_\e \om_\rho{}^{ab} 
  + \frac{2\beta}{\ell} e \bar{\epsilon} \g^{\mu\nu\rho} D_\nu \psi_\rho \Big). \label{eq:svarCSG}
\end{alignat}
Thus the CTMSG is invariant under the local supersymmetry.

By subtracting eq.~(\ref{eq:svarCSG}) from eq.~(\ref{eq:EOMLvarsusy}), 
it is possible to obtain the current conservation for the local supersymmetry.
The supercurrent for the CTMSG is expressed as
\begin{alignat}{3}
  e S^\mu (\epsilon) 
  &= \frac{e}{16\pi G_\text{N}} 
  \Big( \bar{\psi}_\nu \g^{\mu\nu\rho} \dl_\epsilon \psi_\rho 
  + \bar{\epsilon} \g^{\mu\nu\rho} \psi_{\nu\rho} \notag
  \\
  &\quad\,
  - 2\beta \e^{\mu\nu\rho} C_{a\rho} \dl_\epsilon e^a{}_{\nu}
  - \beta \overline{\dl_\e \psi_\nu} \g^{ab} \g^{\mu\nu} D_a \psi_b
  - \frac{2\beta}{\ell} \bar{\epsilon} \g^{\mu\nu\rho} D_\nu \psi_\rho \Big) \notag
  \\
  &=  \frac{1}{16\pi G_\text{N}} \partial_\nu \big( e U^{\mu\nu}(\e) \big). \label{eq:S}
\end{alignat}
Here the antisymmetric tensor $U^{\mu\nu}(\e)$ is given by
\begin{alignat}{3}
  U^{\mu\nu}(\e) &= - 2 \e^{\mu\nu\rho} \bar{\epsilon} \psi_\rho 
  - 2 \beta \bar{\epsilon} \g^{ab} \g^{\mu\nu}  D_a \psi_b. \label{eq:U}
\end{alignat}
In order to derive eq.~(\ref{eq:S}), we used the second line in eq.~(\ref{eq:gammarelation}) 
and imposed the equation of motion $\Psi^\mu = 0$.

%%%%%%%%%%%%%%%%%%%%%%%%%%%%%%%%%%%%%%%%%%%%%%%%%%%%%%%%%%%%%%%%%%%%%%%%%%%%%%%%%%%%%%%%%%%%%%%%%%%
\section{Asymptotic Symmetry Group for AdS$_3$ Geometry}
\label{sec:ASG}
%%%%%%%%%%%%%%%%%%%%%%%%%%%%%%%%%%%%%%%%%%%%%%%%%%%%%%%%%%%%%%%%%%%%%%%%%%%%%%%%%%%%%%%%%%%%%%%%%%%

In this section we briefly review the asymptotic behavior of AdS$_3$ geometry including supersymmetry.
At the spatial infinity $r \to \infty$, the metric of AdS$_3$ geometry becomes
\begin{alignat}{3}
  ds^2 = - N^2 dt^2 + r^2 d\phi^2 + N^{-2} dr^2, \qquad N = \frac{r}{\ell}, \label{eq:AdS3}
\end{alignat}
where $t$, $\phi$ and $r$ are time, angular and radial directions, respectively.
This background corresponds to the massless BTZ black hole.
The Riemann tensor is simply given by 
$R_{\mu\nu\rho\sigma} = - \frac{1}{\ell^2} (g_{\mu\rho} g_{\nu\sigma} - g_{\nu\rho} g_{\mu\sigma})$.
In the background of the massless BTZ black hole, the vielbein and the spin connection become
\begin{alignat}{3}
  &e^0 = \frac{r}{\ell} dt, \qquad e^1 = r d\phi, \qquad e^2 = \frac{\ell}{r} dr, \label{eq:bgviel}
  \\
  &\om_t{}^0{}_2 = \frac{r}{\ell^2}, \qquad \om_\phi{}^1{}_2 = \frac{r}{\ell}. \label{eq:bgspin}
\end{alignat}
$\mu,\nu = t,\phi,r$ are used for spacetime indices and $a,b = 0,1,2$ are done for local Lorentz ones.

Since we are interested in the boundary behavior of the symmetry group,
we explore general coordinate transformation $x'^\mu = x^\mu - \xi^\mu$ 
which does not change the geometry of AdS$_3$ only at the spatial infinity.
The condition to be imposed for the variation of the metric is written as follows.
\ba
  \dl_\xi g_{\mu\nu} = 
  \begin{pmatrix}
    \mc{O}(1) & \mc{O}(1) & \mc{O}(r^{-1}) \\
    \mc{O}(1) & \mc{O}(1) & \mc{O}(r^{-1}) \\
    \mc{O}(r^{-1}) & \mc{O}(r^{-1}) & \mc{O}(r^{-4})
  \end{pmatrix}.
\end{alignat}
The behaviors of the diagonal components of $\dl_\xi g_{\mu\nu}$ are determined so that these go to zero 
faster than the background~(\ref{eq:AdS3}) as $r$ goes to infinity.
Then the behaviors of $\xi^\mu$ and off diagonal components  of $\dl_\xi g_{\mu\nu}$ around the boundary 
are simultaneously fixed.
After some calculations, the general coordinate transformation $\xi^\mu$ 
which satisfy the above condition is solved as
\ba
  \xi^t &= \ell \big( T_+(x^+) + T_-(x^-) \big), \notag
  \\
  \xi^\phi &= T_+(x^+) - T_-(x^-), \label{eq:xiasym}
  \\
  \xi^r &= - r \big( \pa_+ T_+(x^+) + \pa_- T_-(x^-) \big), \notag
\end{alignat}
where $x^\pm = \frac{t}{\ell} \pm \phi$ and $\pa_\pm = \frac{1}{2}(\ell \pa_t \pm \pa_\phi)$.
The isometry group only at the boundary is called asymptotic symmetry group.
The the asymptotic symmetry group is parametrized by arbitrary functions $T_+(x^+)$ and $T_-(x^-)$,
and we often expand these by
\begin{alignat}{3}
  T_{\pm,n}(x^\pm)=\frac{1}{2}e^{inx^\pm}. \label{eq:xiasym2}
\end{alignat}

Now let us calculate the transformation of the vielbein under eq.~(\ref{eq:xiasym}).
As discussed in ref.~\cite{Hyakutake:2012uv}, the transformation should be combined with local Lorentz transformation
$\dl_\Lambda e^a{}_\mu = \Lambda^a{}_b e^b{}_\mu$, where
\begin{alignat}{3}
  \Lambda^a{}_b &= \begin{pmatrix}
    0 & - \pa_+ T_+ + \pa_- T_- & \frac{\ell}{r} \big( \pa_+^2 T_+ + \pa_-^2 T_- \big) \\
    - \pa_+ T_+ + \pa_- T_- & 0 & - \frac{\ell}{r} \big( \pa_+^2 T_+ - \pa_-^2 T_- \big) \\
    \frac{\ell}{r} \big( \pa_+^2 T_+ + \pa_-^2 T_- \big) & \frac{\ell}{r} \big( \pa_+^2 T_+ - \pa_-^2 T_- \big) & 0
  \end{pmatrix}. \label{eq:Lambda}
\end{alignat}
Then the variation $\dl_\xi e^a{}_\mu = \xi^\rho \pa_\rho e^a{}_\mu + \pa_\mu \xi^\rho e^a{}_\rho 
+ \Lambda^a{}_b e^b{}_\mu$ is evaluated as
\begin{alignat}{3}
  \dl_\xi e^a{}_\mu &= \begin{pmatrix}
    0 & 0 & \frac{\ell^2}{r^2} \big( \pa_+^2 T_+ + \pa_-^2 T_- \big) \\
    0 & 0 & - \frac{\ell^2}{r^2} \big( \pa_+^2 T_+ - \pa_-^2 T_- \big) \\
    0 & 0 & 0
  \end{pmatrix}. \label{eq:dlxiviel}
\end{alignat}
This variation goes to zero faster than the background (\ref{eq:bgviel}).
In a similar way, the transformation of the spin connection is given by
$\dl_\xi \om_\mu{}^{ab} = \xi^\rho \pa_\rho \om_\mu{}^{ab} + \pa_\mu \xi^\rho \om_\rho{}^{ab} 
- \pa_\mu \Lambda^{ab} + \Lambda^a{}_c \om_\mu{}^{cb} + \Lambda^b{}_c \om_\mu{}^{ac}$.
After some calculations, the variation of the spin connection becomes
\begin{alignat}{3}
  \dl_\xi \om_t{}^a{}_b &= \begin{pmatrix}
    0 & 0 & - \frac{1}{r} \big( \pa_+^3 T_+ + \pa_-^3 T_- \big) \\
    0 & 0 & \frac{1}{r} \big( \pa_+^3 T_+ - \pa_-^3 T_- \big) \\
    - \frac{1}{r} \big( \pa_+^3 T_+ + \pa_-^3 T_- \big) & - \frac{1}{r} \big( \pa_+^3 T_+ - \pa_-^3 T_- \big) & 0
  \end{pmatrix}, \notag
  \\
  \dl_\xi \om_\phi{}^a{}_b &= \begin{pmatrix}
    0 & 0 & - \frac{\ell}{r} \big( \pa_+^3 T_+ - \pa_-^3 T_- \big) \\
    0 & 0 & \frac{\ell}{r} \big( \pa_+^3 T_+ + \pa_-^3 T_- \big) \\
    - \frac{\ell}{r} \big( \pa_+^3 T_+ - \pa_-^3 T_- \big) & - \frac{\ell}{r} \big( \pa_+^3 T_+ + \pa_-^3 T_- \big) & 0
  \end{pmatrix}, \label{eq:dlxispin}
  \\
  \dl_\xi \om_r{}^a{}_b &= \begin{pmatrix}
    0 & 0 & \frac{\ell}{r^2} \big( \pa_+^2 T_+ + \pa_-^2 T_- \big) \\
    0 & 0 & - \frac{\ell}{r^2} \big( \pa_+^2 T_+ - \pa_-^2 T_- \big) \\
    \frac{\ell}{r^2} \big( \pa_+^2 T_+ + \pa_-^2 T_- \big) & \frac{\ell}{r^2} \big( \pa_+^2 T_+ - \pa_-^2 T_- \big) & 0
  \end{pmatrix}. \notag
\end{alignat}
The variation of the spin connection also goes to zero faster that the background (\ref{eq:bgspin}).
These results will be employed to calculate central charges in the next section.

Next let us explore local supersymmetric transformation $\epsilon(x)$ which satisfy the boundary condition 
at the spatial infinity. Notations are the same as in ref.~\cite{Hyakutake:2012uv}. 
Because $\psi_\mu = 0$ for AdS$_3$ solution, the condition for the supersymmetric variation is imposed as
\ba
  \delta_\epsilon \psi_\mu = 
  \begin{pmatrix}
    \mc{O}(r^{-1/2}) & \mc{O}(r^{-1/2}) & \mc{O}(r^{-5/2}) 
  \end{pmatrix}. \label{eq:Killingspinor}
\end{alignat}
The solution of eq.~(\ref{eq:Killingspinor}) becomes
\begin{alignat}{3}
  \epsilon(x^+) = r^{1/2} \g^0 \chi(x^+) + \ell r^{-1/2} \chi'(x^+), \label{eq:easym}
\end{alignat}
where $\chi(x^+)$ is a Majorana fermion with $\g^2 \chi = \chi$.
The solution depends only on $x^+$, so the remaining local supersymmetry is chiral in this sense.
We often expand $\chi(x^+)$ and $\e(x^+)$ by Fourier modes,
\ba
  \chi_s &= e^{i s x^+} \begin{pmatrix} 0 \\ 1 \end{pmatrix}, \qquad
  \epsilon_s = e^{i s x^+} 
  \begin{pmatrix}
    -r^{1/2} \\ i \ell s r^{-1/2} 
  \end{pmatrix}, \label{eq:easym2}
\end{alignat}
which satisfy the following relation
\begin{alignat}{3}
  \chi_s^T \chi_t = 2 T_{+,s+t}, \qquad
  \overline{\e_s} \g^\mu \e_t &= -2i \xi_{+,s+t}^\mu. \label{eq:normalization}
\end{alignat}
From eq.~(\ref{eq:normalization}), it is clear that $s+t$ should take some integer value.
When $s,t \in \mathbb{Z} + \frac{1}{2}$, those modes are called in the Neveu-Schwarz sector.
On the other hand, when $s,t \in \mathbb{Z}$, those modes are done in the Ramond sector.

%%%%%%%%%%%%%%%%%%%%%%%%%%%%%%%%%%%%%%%%%%%%%%%%%%%%%%%%%%%%%%%%%%%%%%%%%%%%%%%%%%%%%%%%%%%%%%%%%%%
\section{Super Virasoro Algebras from Chiral Supergravity}
\label{sec:superVirasoro}
%%%%%%%%%%%%%%%%%%%%%%%%%%%%%%%%%%%%%%%%%%%%%%%%%%%%%%%%%%%%%%%%%%%%%%%%%%%%%%%%%%%%%%%%%%%%%%%%%%%

So far we have constructed Noether currents for the CTMSG.
Since the CTMSG has stable AdS$_3$ background for the critical point 
$|\beta/\ell|=1$\cite{Li:2008dq,Becker:2009mk}, we consider the chiral supergravity below.
Now we evaluate super Virasoro algebras at the boundary of the chiral supergravity.
The Hamiltonian for the general coordinate transformation $\xi^\mu$ is given by
\ba
  H(\xi) &= \int dr d\phi \, e J^t(\xi) 
  =\frac{1}{16\pi G_\text{N}} \oint_{r=\infty} \!\!\!\!\!\!\!\! d\phi \,
  \big( e Q^{tr}(\xi) + e \tilde{Q}^{tr}(\xi) \big). \label{eq:Ham}
\end{alignat}
The variation of the Hamiltonian is related to the Poisson bracket of the algebra as
\ba
  \dl_{\xi_2} H(\xi_1) = \{H(\xi_1), H(\xi_2) \} = H([\xi_1,\xi_2]) + K(\xi_1,\xi_2).
  \label{eq:bracket1}
\end{alignat}
The last term represents the central extension of the algebra. 
Let us evaluate the above quantity in the background of the massless BTZ black hole
(\ref{eq:AdS3}) with $\psi_\mu = 0$. 
The energy of the massless black hole is zero, so $H(\xi) = 0$ in this background.
Thus $K(\xi_1,\xi_2) = \dl_{\xi_2} H(\xi_1)$ and it is evaluated like
\ba
  \dl_{\xi_2} H(\xi_1) &=\frac{1}{16\pi G_\text{N}} \oint_{r=\infty} \!\!\!\!\!\!\!\! d\phi \, 
  \Big\{ \dl_{\xi_2} \big( e Q^{tr}(\xi_1) \big) 
  + e \big( \xi_1^t \Theta^r(\xi_2) - \xi_1^r \Theta^t(\xi_2) \big) \Big\} \notag
  \\%%%%%%%%%%%%%%%%%%%%%%%%%%%%
  &=\frac{1}{16\pi G_\text{N}} \oint_{r=\infty} \!\!\!\!\!\!\!\! d\phi \, 
  \Big\{ \dl_{\xi_2} \big( 2 e e^t{}_a e^r{}_b \xi_1^\rho \om_\rho{}^{ab} \big) 
  + \frac{\beta}{2} \dl_{\xi_2} \big( 4 C_{\phi \sg} + \om_{\phi ab} \om_\sg{}^{ab} \big) \xi_1^\sg \notag
  \\
  &\qquad\quad\;
  + 4 e \xi_1^{[t} e^{r]}{}_a e^\nu{}_b \dl_{\xi_2} \om_\nu{}^{ab}
  + \beta e \xi_1^{[t} \e^{r]\nu\rho} \big( 4 C_{a\nu} \dl_{\xi_2} e^a{}_\rho 
  + \om_{\nu ab} \dl_{\xi_2} \om_\rho{}^{ab} \big) \Big\} \notag
  \\%%%%%%%%%%%%%%%%%%%%%%%%%%%%
  &= - \frac{\ell}{4\pi G_\text{N}} \oint_{r=\infty} \!\!\!\!\!\!\!\! d\phi \, 
  \Big\{ \Big(1 - \frac{\beta}{\ell} \Big) T_{1+} \pa_+^3 T_{2+} 
  + \Big(1 + \frac{\beta}{\ell} \Big) T_{1-} \pa_-^3 T_{2-} \Big\}. \label{eq:Hcenter}
\end{alignat}
In order to derive the last expression, we used $e \e^{t\phi r} = 1$, eq.~(\ref{eq:dlxiviel}),
eq.~(\ref{eq:dlxispin}) and following relations.
\begin{alignat}{3}
  &\dl_{\xi_2} \big( 4 C_{\rho\sg} + \om_{\rho ab} \om_\sg{}^{ab} \big) = 
  \begin{pmatrix} 
    \frac{4}{\ell^2} (\pa_+^3 T_{2+} + \pa_-^3 T_{2-}) & \frac{4}{\ell} (\pa_+^3 T_{2+} - \pa_-^3 T_{2-}) & 0 \\
    \frac{4}{\ell} (\pa_+^3 T_{2+} - \pa_-^3 T_{2-}) & 4 (\pa_+^3 T_{2+} + \pa_-^3 T_{2-}) & 0 \\
    0 & 0 & 0
  \end{pmatrix}, \notag
  \\%%%%%%%%%%%%%%
  &\dl_{\xi_2} \big( 4 C_{\rho\sg} + \om_{\rho ab} \om_\sg{}^{ab} \big) \xi_1^\sg = 
  \begin{pmatrix} 
    \frac{8}{\ell} (T_{1+} \pa_+^3 T_{2+} + T_{1-} \pa_-^3 T_{2-}) \\
    8 (T_{1+} \pa_+^3 T_{2+} - T_{1-} \pa_-^3 T_{2-}) \\
    0
  \end{pmatrix},
  \\%%%%%%%%%%%%%%%
  &4 C_{a\nu} \dl_{\xi_2} e^a{}_\rho + \om_{\nu ab} \dl_{\xi_2} \om_\rho{}^{ab} = 
  \begin{pmatrix} 
    \frac{2}{\ell^2} (\pa_+^3 T_{2+} + \pa_-^3 T_{2-}) & \frac{2}{\ell} (\pa_+^3 T_{2+} - \pa_-^3 T_{2-}) & 0 \\
    \frac{2}{\ell} (\pa_+^3 T_{2+} - \pa_-^3 T_{2-}) & 2 (\pa_+^3 T_{2+} + \pa_-^3 T_{2-}) & 0 \\
    0 & 0 & 0
  \end{pmatrix}. \notag
\end{alignat}
Notice that left and right modes are separated in a nontrivial way in eq.~(\ref{eq:Hcenter}).

Now we substitute the Fourier mode expansion of eq.~(\ref{eq:xiasym2}). Then the variation
of the Hamiltonian becomes
\ba
  \dl_{\xi_{\pm,n}} H(\xi_{\pm,m}) &= 
  - i \frac{\ell}{8 G_\text{N}} \Big(1 \mp \frac{\beta}{\ell} \Big) m^3 \dl_{m+n,0}.
\end{alignat}
This gives the central extensions of left and right Virasoro algebras.
By expanding $H(\xi_{\pm,m}) = L^\pm_m e^{imx^\pm}$ and replacing the Poisson bracket with the commutator, 
we obtain Virasoro algebras for left and right modes.
\begin{alignat}{3}
  [L^+_m, L^+_n] &= (m-n) L^+_{m+n} + \frac{c_+}{12} m^3 \dl_{m+n,0}, \notag
  \\
  [L^-_m, L^-_n] &= (m-n) L^-_{m+n} + \frac{c_-}{12} m^3 \dl_{m+n,0}. \label{eq:Virasoroalg}
\end{alignat}
Here the central charges are given by
\begin{alignat}{3}
  c_\pm = \frac{3\ell}{2G_\text{N}} \Big( 1 \mp \frac{\beta}{\ell} \Big). \label{eq:cpm}
\end{alignat}
Note that the sign is flipped compared with ref.~\cite{Hotta:2008yq} 
because of the definition $e \e^{t\phi r} = 1$.
At the critical point, one of the central charges vanishes.

Next let us evaluate the Poisson bracket of the supercharge.
The supercharge for the local supersymmetry is written as
\ba
  F(\e) &= \int drd\phi \, e S^t(\e) 
  = \frac{1}{16\pi G_\text{N}} \oint_{r=\infty} \!\!\!\!\!\!\!\! d\phi \, e U^{tr}(\e). \label{eq:supercharge}
\end{alignat}
It is obvious that the supercharge is zero in the background of $\psi_\mu = 0$.
The variation of the supercurrent under the local supersymmetry is evaluated as
\ba
  \delta_{\e_2} F(\e_1) &= \{ F(\e_1), F(\e_2) \} 
  = H(\overline{\e_1}\g \e_2) + K(\e_1,\e_2), \label{eq:Fcenter}
\end{alignat}
where $K(\e_1,\e_2)$ is the central extension of the algebra.
Let us evaluate the above quantity in the background of the massless BTZ black hole
(\ref{eq:AdS3}) with $\psi_\mu = 0$. 
The energy of the massless black hole is zero, so $H(\xi) = 0$ in this background.
Thus $K(\e_1,\e_2) = \dl_{\e_2} F(\e_1)$ and its explicit form is calculated as
\begin{alignat}{3}
  \delta_{\epsilon_2} F(\epsilon_1) 
  &= \frac{1}{16\pi G_\text{N}} \oint_{r=\infty} \!\!\!\!\!\!\!\! d\phi \,
  \Big( 4 \overline{\epsilon_1} \mathcal{D}_\phi \epsilon_2
  - \frac{\beta}{2} e R_{\rho\sigma ab} \overline{\epsilon_1} \g^{ab} \g^{tr} \g^{\rho\sigma} \epsilon_2
  - \frac{2\beta}{\ell} e \overline{\epsilon_1} \g^{ab} \g^{tr} \g_b D_a \epsilon_2 \Big) \notag
  \\%%%%%%%%%%%%%%%%%%%%
  &= \frac{1}{4\pi G_\text{N}} \Big( 1 - \frac{\beta}{\ell} \Big) \oint_{r=\infty} \!\!\!\!\!\!\!\! d\phi \,
  \overline{\epsilon_1} \mathcal{D}_\phi \epsilon_2, \notag
  \\%%%%%%%%%%%%%%%%%%%%
  &= \frac{i \ell}{4\pi G_\text{N}} \Big( 1 - \frac{\beta}{\ell} \Big) \oint_{r=\infty} \!\!\!\!\!\!\!\! d\phi \,
  \chi_1^T \chi''_2.
\end{alignat}
In the above we employed eq.~(\ref{eq:gammarelation}).
Let us substitute Fourier mode expansion of eq.~(\ref{eq:easym2}). 
Then the variation of the supercharge is evaluated as
\ba
  \delta_{\e_t} F(\e_s) &= -i \frac{\ell}{2G_\text{N}} \Big( 1 - \frac{\beta}{\ell} \Big) s^2 \dl_{s+t,0}. 
  \label{eq:Fcenter2}
\end{alignat}
This corresponds to the central extension of the super Virasoro algebra.
Notice that $i\overline{\e_s} \g^\mu \e_t = 2 \xi_{+,s+t}^\mu$. 
By expanding $F(\e_s) = G_s e^{isx^+}$, the algebra is expressed as
\begin{alignat}{3}
  \{ G_s, G_t \} &= 2 L_{s+t} + \frac{c_+}{3} s^2 \dl_{s+t,0}. \label{eq:superV}
\end{alignat}
The Neveu-Schwarz sector corresponds to $s,t \in \mathbb{Z} + \frac{1}{2}$, and
the Ramond sector does to $s,t \in \mathbb{Z}$.

Finally let us examine the variation of the supercharge under the general coordinate 
transformation\cite{Hyakutake:2012uv}.
When the transformation $\xi^\mu_+$ depends only on $x^+$, we obtain
\begin{alignat}{3}
  \dl_{\xi_+} F(\e_1) 
  &= \{ F(\e_1), H(\xi_+) \} 
  = - F(\dl_{\xi_+} \e_1),
\end{alignat}
where $\dl_{\xi_+} \e_1 = \xi_+^\rho \pa_\rho \e_1 + \frac{1}{4} \Lambda_{ab} \g^{ab} \e_1$.
Notice that the integral constant should be zero since $F(\e) = 0$ for $\psi_\mu = 0$.
By setting $\xi_+ = \xi_{+,m}$ and $\e_1 = \e_s$, we obtain
\begin{alignat}{3}
  [ L_m^+, G_s ] = \Big(\frac{m}{2} - s \Big) G_{m+s}.
\end{alignat}
In a similar way, it is possible to show $[ L_m^-, G_s ] = 0$.
Therefore we conclude that there are left and right Virasoro algebras at the boundary 
with different central charges, and left mode is extended to the super Virasoro algebra.

%%%%%%%%%%%%%%%%%%%%%%%%%%%%%%%%%%%%%%%%%%%%%%%%%%%%%%%%%%%%%%%%%%%%%%%%%%%%%%%%%%%%%%%%%%%%%%%%%%%%%%%

\section{Conclusion and Discussion}

%%%%%%%%%%%%%%%%%%%%%%%%%%%%%%%%%%%%%%%%%%%%%%%%%%%%%%%%%%%%%%%%%%%%%%%%%%%%%%%%%%%%%%%%%%%%%%%%%%%%%%%

In this note, we investigated the chiral supergravity in three dimensions.
The charges for the general coordinate transformation and local supersymmetry
are explicitly constructed by applying Wald's prescription to the vielbein formalism.
Commutation relations of the charges are explored in detail and super Virasoro algebras are constructed
for AdS$_3$ background.
Especially, the central extensions of the left and right super Virasoro algebras are evaluated 
by calculating the variations of the charges. The asymmetric central charges are obtained 
and those expressions are given by $c_\pm = \frac{3\ell}{2G_\text{N}} (1 \mp \frac{\beta}{\ell})$.

Note that the super Virasoro algebras (\ref{eq:Virasoroalg}) and (\ref{eq:superV}) are not in the canonical form.
In order to make the expressions canonical, we just shift the zero point energy as
\begin{alignat}{3}
  L_0^\pm \;\to\; L_0^\pm - \frac{c_\pm}{24}.
\end{alignat}
Then the algebras become
\begin{alignat}{3}
  [L^\pm_m, L^\pm_n] &= (m-n) L^\pm_{m+n} + \frac{c_\pm}{12} (m^3-m) \dl_{m+n,0}, \notag
  \\
  \{ G_s, G_t \} &= 2 L_{s+t} + \frac{c_+}{3} \Big(s^2 - \frac{1}{4}\Big) \dl_{s+t,0}. \label{eq:sV}
\end{alignat}
At the same time, the energy of the global AdS$_3$ geometry is shifted to zero.
Thus the effective central charge is the same as the central charge, and
the entropy of the BTZ black hole can be correctly explained by 
the Cardy formula\footnote{As a review on the BTZ black hole entropy and Cardy formula, 
see ref.~\cite{Carlip:1998qw} for example.}.
Though this conclusion was obtained in the supersymmetric theory, it is also true
for the bosonic case if we truncate the fermionic sector.

Since the vielbein formulation of the chiral supergravity is well established,
it is interesting to apply these results to other geometries, 
such as warped AdS$_3$\cite{Anninos:2008fx,Compere:2008cv}, or
Kerr/CFT correspondence\cite{Guica:2008mu}.
For these cases, it is important to generalize the covariant formalism of 
refs.~\cite{Barnich:2001jy,Barnich:2007bf} to the chiral supergravity.
It is also important to apply the vielbein formalism to the higher spin supergravity
and derive the central charges\cite{Creutzig:2011fe,Henneaux:2012ny,Hanaki:2012yf,Gaberdiel:2014yla}.

\section*{Acknowledgement}

The author would like to thank Yuji Sugawara and Takahiro Nishinaka.
This work was partially supported by the Ministry of Education, Science, 
Sports and Culture, Grant-in-Aid for Young Scientists (B) 24740140, 2012.

%%%%%%%%%%%%%%%%%%%%%%%%%%%%%%%%%%%%%
%\vfill\eject

\end{document}